 \documentclass[authorversion]{acmart}

\usepackage{graphicx}
\usepackage{subcaption}
\usepackage{amsfonts}
\usepackage{amsmath}
\usepackage{booktabs}
\usepackage{enumitem}
\usepackage{soul}

\usepackage{blindtext}

\usepackage{xpatch}

\makeatletter
\xpatchcmd{\ps@firstpagestyle}{Manuscript submitted to ACM}{}{\typeout{First patch succeeded}}{\typeout{first patch failed}}
\xpatchcmd{\ps@standardpagestyle}{Manuscript submitted to ACM}{}{\typeout{Second patch succeeded}}{\typeout{Second patch failed}}    \@ACM@manuscriptfalse% Also in titlepage
\makeatother
\renewcommand\footnotetextcopyrightpermission[1]{} % removes footnote with conference info
\setcopyright{none}
\title{Interactive Sound Rendering on Mobile Devices using Ray-Parameterized Reverberation Filters}

\author{Carl Schissler, Dinesh Manocha}
\affiliation{%
\institution{University of North Carolina at Chapel Hill}
}


\begin{document}

%%% A ``teaser'' image appears under the title and affiliation information,
%%% horizontally centered, and above the two columns of text. This is OPTIONAL.
%%% If you choose to have a ``teaser'' image, it needs to be placed between
%%% ``\begin{document}'' and ``\maketitle.''

\begin{teaserfigure}
	\centering
	%add desired spacing between images, e. g. ~, \quad, \qquad etc.
   	\begin{subfigure}[b]{0.32\textwidth}
	       \centering
	       \includegraphics[height=1.31in]{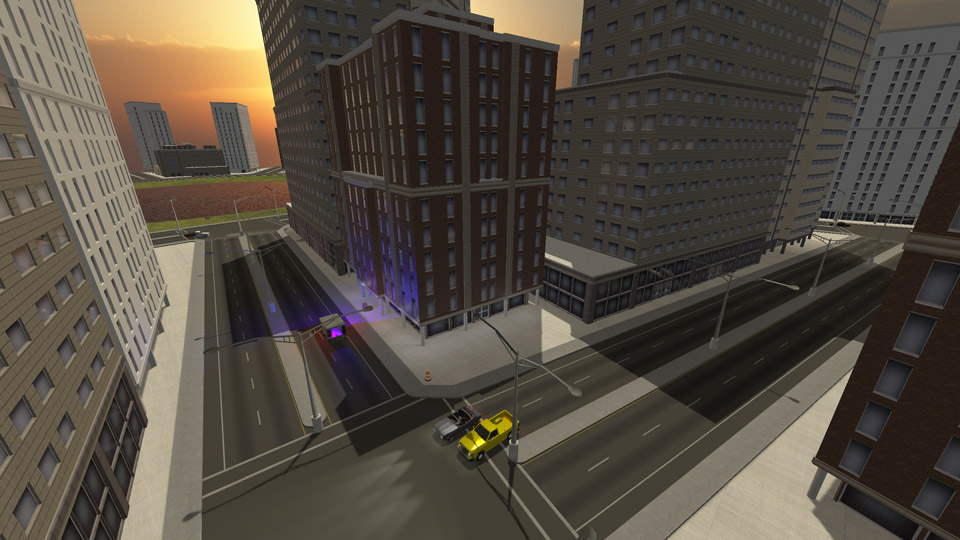}
	\end{subfigure}
	~
	\begin{subfigure}[b]{0.32\textwidth}
	       \centering
	       \includegraphics[height=1.31in]{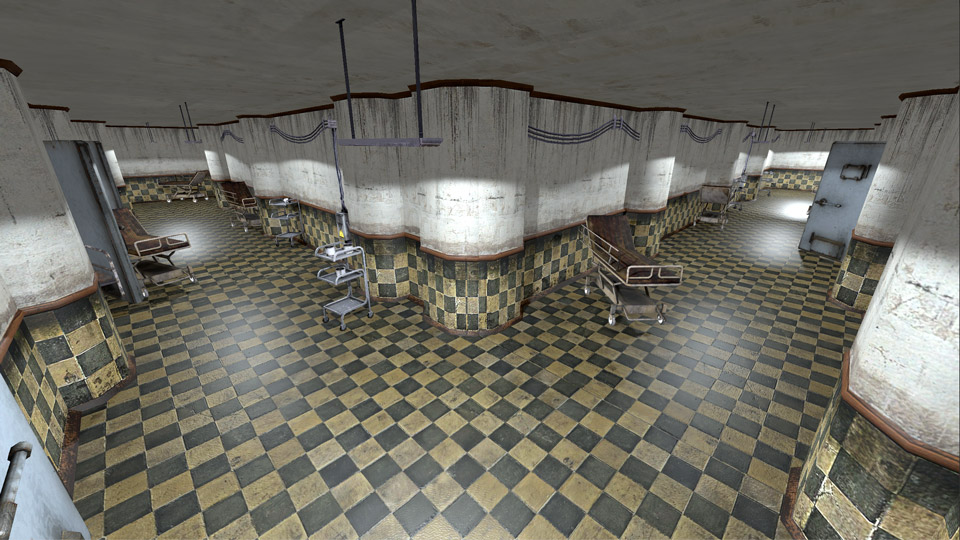}
		\end{subfigure}
	~
   	\begin{subfigure}[b]{0.32\textwidth}
	       \centering
	       \includegraphics[height=1.31in]{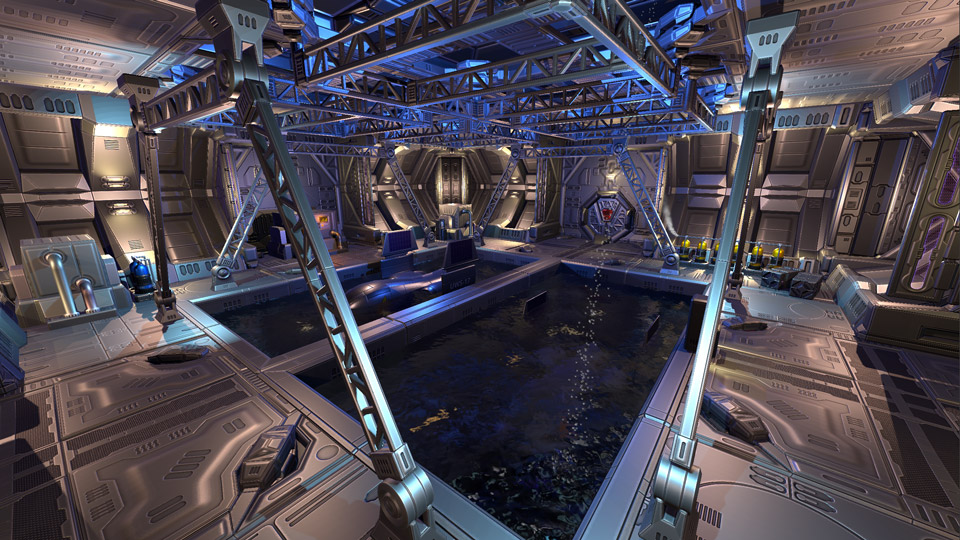}
	\end{subfigure}
	\caption{
Our sound rendering approach can compute realistic audio on desktop and mobile for complex dynamic scenes: City (L), Hospital (C), Sub Bay (R).
} \label{fig:teaser}
\end{teaserfigure}

%%% Your paper's abstract goes in its own section.
\begin{abstract}
We present a new sound rendering pipeline that is able to generate plausible sound propagation effects for interactive dynamic scenes.
Our approach combines ray-tracing-based sound propagation with reverberation filters using robust automatic reverb parameter estimation that is driven by impulse responses computed at a low sampling rate.
We propose a unified spherical harmonic representation of directional sound in both the propagation and auralization modules and use this formulation to perform a constant number of convolution operations for any number of sound sources while rendering spatial audio.
In comparison to previous geometric acoustic methods, we achieve a speedup of over an order of magnitude while delivering similar audio to high-quality convolution rendering algorithms.
As a result, our approach is the first capable of rendering plausible dynamic sound propagation effects on commodity smartphones.
\end{abstract}

\maketitle

% <concept_id>10010147.10010371.10010382</concept_id>
% <concept_desc>Computing methodologies~Image manipulation</concept_desc>
% <concept_id>10010147.10010371.10010382.10010236</concept_id>
% <concept_desc>Computing methodologies~Computational photography</concept_desc>
% <concept_significance>300</concept_significance>



% The next three commands are required, and insert the user-generated keywords, 
% The CCS concepts list, and the rights management text.
% Please make sure there is a blank line between each of these three commands.

% \keywordlist

% \conceptlist

% \printcopyright

\section{Introduction}
\label{sec:intro}

Sound rendering is frequently used to increase the sense of realism in 
virtual reality (VR) and augmented reality (AR) applications.
A recent trend has been to use mobile devices such as Samsung Gear VR\texttrademark and Google Daydream-ready phones\texttrademark for VR.
A key challenge is to generate realistic sound propagation effects in dynamic scenes on these low-power devices.

%However, a significant challenge is rendering high-quality 3D audio-visual content with low latency on hardware that has significantly less computational power.
%While there is much work on optimizing visual rendering for slow hardware using level-of-detail approaches, and precomputed lighting, there is comparatively little research that focuses on physically-based sound rendering for low-power devices.
%Dynamic scenes with moving sources, listeners, and objects present an additional challenge because precomputed filters cannot be used.

A major component of rendering plausible sound is the simulation of sound propagation within the virtual environment.
When sound is emitted from a source it travels through the environment and may undergo reflection, diffraction, scattering, and transmission effects before the sound is heard by a listener.

The most accurate interactive techniques for sound propagation and rendering are based on a {\em convolution-based sound rendering} pipeline
that splits the computation into three main components.
The first, the sound propagation module, uses geometric algorithms like ray or beam tracing to simulate how sound travels through the environment and computes an impulse response (IR) between each source and listener.
The second takes the IR and converts it into a spatial impulse response (SIR) that is suitable for auralization of directional sound.
Finally, the auralization module convolves each channel of the SIR with the anechoic audio for the sound source to generate the audio which is reproduced to the listener through an auditory display device (e.g. headphones).

Current algorithms that use a convolution-based pipeline can generate high-quality interactive audio for scenes with dozens of sound sources on commmodity desktop or laptop machines~\cite{lentz2007,savioja2015,schissler2016a}.
However, these methods are less suitable for low-power mobile devices where there are significant computational and memory constraints.
The IR contains directional and frequency-dependent data at the audio rendering sample rate (e.g. $44.1$kHz) and therefore can require up to $10-15$MB per sound source, depending on the number of frequency bands, length of the impulse response, and the directional representation.
%The auralization module requires double or triple-buffering the SIR so that it can update the impulse response with interpolation and low latency.
%This adds another few MB per sound source to the memory requirements.
%In total, the memory required is about 10-15MB per sound source when the impulse responses are 2 seconds long.
This large memory usage severely constrains the number of sources that can be simulated concurrently.
In addition, the number of rays that must be traced during sound propagation to avoid an aliased or noisy IR can be large and take $100$ms to compute on a multi-core desktop CPU for complex scenes.
%Even on a fast desktop computer, it may take 100ms to compute a single frame for complex scenes~\cite{schissler2016a}.
The construction of the SIR from the IR is also an expensive operation that takes about 20-30ms per source for a single desktop CPU thread~\cite{schissler2017b}.
Convolution with the SIR requires time proportional to the length of the impulse response, and the number of concurrent convolutions is limited by the tight real-time deadlines needed for smooth audio rendering without clicks or pops.

A low-cost alternative to convolution-based sound rendering is to use {\em artificial reverberators}.
Artificial reverberation algorithms use recursive feedback-delay networks to simulate the decay of sound in rooms~\cite{gardner2002}.
These filters are typically specified using different parameters like the reverberation time, direct-to-reverberant sound ratio, predelay, reflection density, directional loudness, etc.
These parameters are either specified by an artist or approximated using scene characteristics~\cite{tsingos2009precomputing,antani2013aural}.
However, most prior approaches for rendering artificial reverberation assume that the reverberant sound field is completely diffuse.
As a result, they cannot be used to efficiently generate accurate directional reverberation or time-varying effects in dynamic scenes. 
%such as with outdoor scenes or a sound source at the end of a long hallway.
Compared to convolution-based rendering, previous artificial reverberation methods suffer from reduced quality of spatial sound and can have difficulties in automatic determination of dynamic reverberation parameters.

{\bf Main Results:}
We present a new approach for sound rendering that combines ray-tracing-based sound propagation with reverberation filters to generate smooth, plausible audio for dynamic scenes with moving sources and objects.
The main idea of our approach is to dynamically compute reverberation parameters using an interactive ray tracing algorithm that computes an IR with a low sample rate.
Moreover, direct sound, early reflections, and late reverberation are rendered using the spherical harmonic basis functions, and this allows our approach to capture many important features of the impulse response, including the directional effects.
The number of convolution operations performed in our integrated pipeline is constant, as this computation is performed only for the listener
and does not scale with the number of sources.
Moreover, we perform convolutions with very short impulse responses for spatial sound.
%when compared to previous convolution-based approaches.
%As compared to convolution-based sound rendering approaches, our integrated method consumes significantly less memory and CPU resources.
%
%Our approach is able to reduce the number of rays traced during sound propagation by about $10$X compared to convolution-based rendering, and the auralization cost is reduced by $2-3$X and is invariant with the impulse response length.
%
%Some of the novel components of our approach include:
%\begin{itemize}
%\item Ray-parameterized reverberation which can automatically and robustly compute the reverberation parameters in dynamic scenes.
%\item Our approach uses an order of magnitude fewer rays than previous interactive geometric propagation methods.
%\item We use a unified representation based on spherical harmonics for propagation as well as auralization. 
%This allows our method to reproduce various characteristics of the impulse response, including directional effects.
%Moreover, we use a constant of convolution operations for sound rendering. 
%\item For the first time, we demonstrate an approach that can render high-quality sound propagation on a mobile device with both low memory and computational overhead.
%\end{itemize}
We have both quantitatively and subjectively evaluated our approach on various interactive scenes with $7-23$ sources and observe significant improvements of $9-15$x compared to convolution-based sound rendering approaches.
Furthermore, our technique reduces the memory overhead by about $10$x.
For the first time, we demonstrate an approach that can render high-quality interactive sound propagation on a mobile device with both low memory and computational overhead.

\section{Background} \label{sec:background}
In this section, we give a brief overview of prior work on sound propagation, auralization and spatial sound.
%%%%%%%%%%%%%%%%%%%%%%%%%%%%%%%%%%%%%%%%%%%%%%%%%%%%%%%%%%%%%%%%%%%%%%%%%%%%%%%%

{\bf Sound Propagation:}
Methods for computing sound propagation and impulse responses in virtual environments can be divided into two broad categories: wave-based sound propagation and geometric sound propagation.
Wave-based sound propagation techniques directly solve the acoustic wave equation in either time domain~\cite{savioja2010,raghuvanshi2014} or frequency domain~\cite{ciskowski1991,mehra2013} using numerical methods. They are the most accurate methods, but scale poorly with the size of the domain and the maximum frequency. Current precomputation-based wave propagation methods are limited to static scenes.
%Time-domain wave solvers generally involve dividing the simulation domain into volume elements and calculating how sound pressure propagates within and between the volume elements.
%These include FDTD~\cite{savioja2010} and adaptive rectangular decomposition~\cite{raghuvanshi2009}.
%These methods directly compute a pressure impulse response between each source and listener position in the scene.
%Frequency-domain wave solvers instead solve for the complex pressure at each frequency, and include the boundary element method~\cite{ciskowski1991} and equivalent source method~\cite{mehra2013}.
%While wave-based sound propagation techniques are accurate, 
%As a result, they are rarely used for interactive applications, except as a precomputation for static scenes~\cite{}.
Geometric sound propagation techniques make the simplifying assumption that surface primitives are much larger than the wavelength of sound~\cite{savioja2015}.
As a result, they are better suited for interactive applications, but do not inherently simulate low-frequency diffraction effects. Some techniques based on Uniform theory of diffraction have been used to approximate diffraction effects for interactive applications~\cite{tsingos2001,schissler2014}
%These include the approximate uniform theory of diffraction (UTD)~\cite{kouyoumjian1974} and the more accurate Biot-Tolstoy-Medwin (BTM) method~\cite{svensson1999}.
%Diffraction may also be computed probabalistically using a technique based on the Heisenberg uncertainty principle~\cite{stephenson2010}.
%For the computation of other sound phenomena such as specular and diffuse reflections, various algorithms have been proposed.
Specular reflections are frequently computed using the image source method (ISM), which can be accelerated using ray tracing~\cite{vorlander1989} or beam tracing~\cite{funkhouser1998}.
The most common techniques for diffuse reflections are based on Monte Carlo path or sound particle tracing~\cite{vorlander1989,embrechts2000}.
%In these methods, many rays are emitted from the source or listener that each carry some sound energy.
%These rays are then propagated through the environment until a path to a source or listener is detected that contributes to the impulse response.
Ray tracing may be performed from either the source, listener, or from both directions~\cite{cao2016} and can be improved by utilizing temporal coherence~\cite{schissler2014}. Our approach can be combined with any ray-tracing based interactive sound propagation algorithm.
%The performance of path tracing can be improved through the use of temporal coherence techniques that use sound propagation results from previous frames~\cite{schissler2014,schissler2016a}.
%Radiosity has also been proposed as a way to compute diffuse reflections, though it is limited to static scenes~\cite{nosal2004}.

%%%%%%%%%%%%%%%%%%%%%%%%%%%%%%%%%%%%%%%%%%%%%%%%%%%%%%%%%%%%%%%%%%%%%%%%%%%%%%%%

% Convolution
% Fractional delay interpolation
% Artificial reverb
{\bf Auralization:}
In convolution-based sound rendering, an impulse response (IR)  must be convolved with the dry source audio.
The fastest convolution techniques are based on convolution in the frequency domain.
To achieve low latency, the IR is typically partitioned into blocks with smaller partitions toward the start of the IR~\cite{gardner1994}.
Time-varying IRs can be handled by rendering two convolution streams simultaneously and interpolating between their outputs in the time domain~\cite{muller2001}.
Artificial reverberation methods approximate the reverberant decay of sound energy in rooms using recursive filters and feedback delay networks~\cite{gardner2002,valimaki2012}.
One of the earliest and most widely used reverberator designs was proposed by Schroeder~\shortcite{schroeder1961}.
%Its quality can be improved by increasing the number of comb and all pass filters, or by inserting low-pass filters in the comb filter feedback paths to simulate air absorption~\cite{moorer1979}.
Artificial reverberation has also been extended to B-format ambisonics~\cite{anderson2009}.
%However, these methods can't reproduce time-varying acoustic effects.
%These techniques are more efficient and their complexity doesn't depend on the length of the impulse response length.
%However, they cannot easiy reproduce all of the directional and time-varying effects rendered by convolution-based pipeline, and it is difficult to robustly determine the reverberation parameters for a general impulse response.

%MOVE THIS TO SECTION 3 or 4 WHERE YOU USE IT: 

%We extend these ideas in our work to better control the reverb directivity.

%Fractional delay interpolation has been proposed as an efficient method for rendering Doppler effects~\cite{strauss1998}.
%This technique involves reading from a circular buffer of source audio samples at a delayed offset, where the rate of interpolation depends on the relative velocity along the propagation path.
%Typically, delay interpolation is used to render the direct sound and early reflections, while convolution can be used for the rest of the sound propagation paths~\cite{schissler2016c}.
%Perceptual clustering and culling techniques have also been proposed to reduce the number of propagation paths and sources that must be rendered~\cite{tsingos2004,moeck2007}.

%%%%%%%%%%%%%%%%%%%%%%%%%%%%%%%%%%%%%%%%%%%%%%%%%%%%%%%%%%%%%%%%%%%%%%%%%%%%%%%%

% VBAP
% HRTF
% Spatial sound for propagation
{\bf Spatial Sound:}
In spatial sound rendering, the goal is to reproduce directional audio that gives the listener a sense that the sound is localized in 3D space.
This involves modeling the impacts of the listener's head and torso on the sound received at each ear.
%including inter-aural time and level differences (ITD and ILD) as well as frequency filtering.
The most computationally efficient methods are based on vector-based amplitude panning (VBAP)~\cite{pulkki1997}, which
 compute the amplitude for each channel based on the direction of the sound source relative to the nearest speakers and are suited for reproduction on surround-sound systems.
%However, VBAP doesn't model any ITD or filtering effects and so is not suitable for use with headphones.
Head-related transfer functions (HRTFs) are widely used to model spatial sound that can incorporate all spatial sound phenomena using measured IRs on a spherical grid surrounding the listener~\cite{moller1992}.
%The HRTF, represented by functions $H_L(\vec{x},t)$ and $H_R(\vec{x},t)$ for the left and right ears, is convolved with the incoming sound source audio to generate the audio at the entrance to the listener's ear canals.

% prior work with the HRTF is focused on point sources, recent work has 
%been proposed to handle area sound sources~\cite{schissler2016b}.
%The HRTF can also be efficiently applied to the sound propagation impulse response in a perceptually-based manner~\cite{schissler2017b}.

%%%%%%%%%%%%%%%%%%%%%%%%%%%%%%%%%%%%%%%%%%%%%%%%%%%%%%%%%%%%%%%%%%%%%%%%%%%%%%%%

{\bf Spherical Harmonics:}
Our approach uses spherical harmonic (SH) basis functions.
SH are a set of orthonormal basis functions $Y_{lm}(\vec{x})$ defined on the spherical domain $\mathbb{S}$, where $\vec{x}$ is a vector of unit length, $l = 0, 1, ... n$ and $m = -l, ... ,0, ... l$, and $n$ is the spherical harmonic order.
For SH order $n$, there are $(n+1)^2$ basis functions.
Due to their orthonormality, SH basis function coefficients can be efficiently rotated using an $(n+1)^2$ by $(n+1)^2$ block-diagonal matrix~\cite{ivanic1996}.
While the SH are defined in terms of spherical coordinates, they can be evaluated for Cartesian vector arguments using the fast formulation of~\cite{sloan2013} that uses constant propagation and branchless code to speed up the function evaluation.
SHs have been used as a representation of spherical data such as the HRTF~\cite{rafaely2010,romigh2015}, and also form the basis for the ambisonic spatial audio technique~\cite{gerzon1973}.

\section{Overview} \label{sec:overview}

%%%%%%%%%%%%%%%%%%%%%%%%%%%%%%%%%%%%%%%%%%%%%%%%%%%%%%%%%%%%%%%%%%%%%%%%%%%%%%%%

% Introduction / motivation
The most accurate sound rendering algorithms are based on a convolution-based sound rendering pipeline.
%of the spatial impulse response with the sound source audio.
However, low-latency convolution is computationally expensive, and so these approaches are limited in terms of number of simultaneous sources they can render~\cite{lentz2007}.
The convolution cost also increases considerably for long impulse responses that are computed in reverberant environments.
As a result, convolution-based rendering pipelines are not practical on current low-power mobile devices.

\begin{figure}
\centering
\includegraphics[width=0.96\columnwidth]{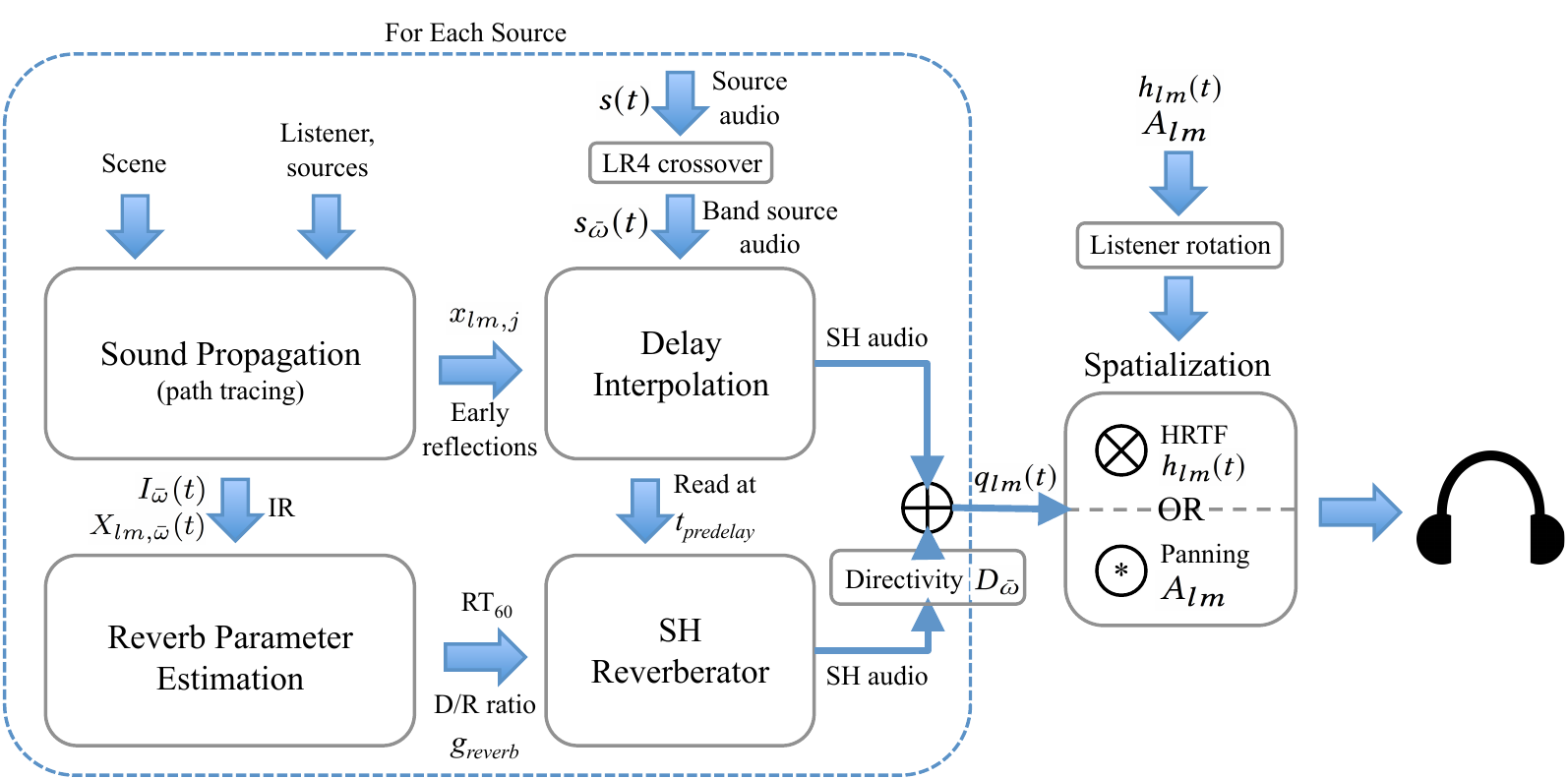}
\caption{Our integrated sound propagation and auralization pipeline. We highlight different components of the new pipeline.
}\label{fig:pipeline_new}
\end{figure}

% Introduce technique at high level
%\subsection{Integrated Propagation \& Auralization}
We present a new integrated approach for sound rendering that performs propagation and spatial sound auralization using ray-parameterized reverberation filters.
Our goal is generate high-quality spatial sound for direct sound, early reflections, and directional late reverberation with significantly less computational overhead than convolution-based techniques.
Our approach renders audio in the spherical harmonic (SH) domain and facilitates spatialization with the user's head-related transfer function (HRTF) or amplitude panning.
%than traditional techniques.
An overview of this pipeline is shown in Figure~\ref{fig:pipeline_new}.
The sound propagation module uses ray and path tracing to estimate the directional and frequency-dependent IR at a low sampling rate (e.g. $100$Hz).
From this IR, we robustly estimate the reverberation parameters, such as the reverberation time (RT$_{60}$) and direct-to-reverberant sound ratio (D/R) for each frequency band.
This information is used to parameterize the artificial reverberator.
Due to the robustness of our parameter estimation and auralization algorithm, our approach is able to use an order of magnitude fewer rays than convolution-based rendering in the sound propagation module.
The artificial reverberator renders a separate channel for each frequency band and SH coefficient, and uses spherical harmonic rotations in the comb-filter feedback path to mix the SH coefficients and produce a natural distribution of directivity for the reverberation decay.
At the reverberation output, we apply frequency-dependent directional loudness to the reverberation signal in order to model the overall frequency-dependent directivity and then sum the audio into a broadband signal in the SH domain.
For the direct sound and early reflection, monaural samples are interpolated from a circular delay buffer of dry source audio and are multiplied by the reflection's SH coefficients.
The resulting audio for the early reflections are mixed with the late reverberation in the SH domain.
This audio is computed for every sound source and then mixed together.
Then in a final spatialization step, the audio for all sources is convolved with a rotated version of the user's HRTF in the SH domain.
The result is spatialized direct sound, early reflections and late reverberation with the directivity information.

\begin{table}[t]
\small
	\centering
	\begin{tabular}{ c|c } 
	\toprule
	\bf{Symbols} & \bf{Meaning} \\
	\midrule
	$n$ & Spherical harmonic order \\
	$N_{\bar{\omega}}$ & Frequency band count \\
	$\bar{\omega}$ & Frequency band \\
	$\vec{x}$ & Direction toward source along propagation path \\
	$x_{lm,j}$ & SH Distribution of sound for $j$th path \\
	$X(\vec{x},t)$ & Distribution of incoming sound at listener in the IR \\
	$X_{lm}(t)$ & Spherical harmonic projection of $X(\vec{x},t)$ \\
	$X_{lm,\bar{\omega}}(t)$ & $X_{lm}(t)$ for frequency band $\bar{\omega}$ \\
	$I_{\bar{\omega}}(t)$ & IR in intensity domain for band $\bar{\omega}$ \\
	\midrule
	$s(t)$ & Anechoic audio emitted by source \\
	$s_{\bar{\omega}}(t)$ & Source audio filtered into frequency bands $\bar{\omega}$ \\
	$q_{lm}(t)$ & Audio at listener position in SH domain \\
	$H(\vec{x},t)$ & Head-related transfer function \\
	$h_{lm}(t)$ & HRTF projected into SH domain \\
	$A(\vec{x})$ & Amplitud panning function \\
	$A_{lm}$ & Amplitude panning function in SH domain \\
	$\mathcal{J}(\mathcal{R})$ & SH rotation matrix for $3 \times 3$ matrix $\mathcal{R}$ \\
	$\mathcal{R}_L$ & $3 \times 3$ matrix for listener head orientation \\
	\midrule
	RT$_{60}$ & Time for reverberation to decay by $60$dB \\
	$g_{comb}^i$ & Feedback gain for $i$th recursive comb filter \\
	$t_{comb}^i$ & Delay time for $i$th recursive comb filter \\
	$g_{reverb,\bar{\omega}}$ & Output gain of SH reverberator for band $\bar{\omega}$ \\
	$t_{predelay}$ & Time delay of reverb relative to $t=0$ in IR \\
	$D_{\bar{\omega}}$ & SH directional loudness matrix \\
	$\tau$ & Temporal coherence smoothing time (seconds) \\
	\bottomrule
	\end{tabular}
\caption{
A table of mathematical symbols used in the paper.}
\label{table:notations}
\vspace*{-0.15in}
\end{table}

\section{Ray-Parameterized Reverberation Filters}
\label{sec:reverb}

%%%%%%%%%%%%%%%%%%%%%%%%%%%%%%%%%%%%%%%%%%%%%%%%%%%%%%%%%%%%%%%%%%%%%%%%%%%%%%%%

% Describe impulse response structure
The goal of our approach is to render artificial reverberation that closely matches the audio generated by convolution-based techniques.
As a result, it is important to replicate the directional frequency-dependent time-varying structure of a typical IR, including direct sound, early reflections (ER), and late reverberation (LR).
%The IR can be divided into 3 main components: direct sound, early reflections (ER), and late reverberation (LR).
%These are illustrated in Figure~\ref{fig:ir_parts}.
%The direct sound is the first sound arrival at the listener's position and is strongly directional.
%The early reflections (ER) follow the direct sound and consist of the first few indirect bounces of sound.
%Early reflections usually arrive from several distinct directions corresponding to major reflectors in the scene.
%The remainder of the impulse response consists of late reverberation (LR).
%The LR represents the buildup of many high-order reflections as the sound decays to zero amplitude.
%Usually, the earlier parts of the LR tends to be directional, then increasingly diffuse towards the later parts as the reflections scatter the sound.
%The rate of decay of the LR varies with frequency and is determined by the effects of materials and air absorption.
%In outdoor environments, the impact of LR is less because the open environment allows much of the later indirect sound to escape the scene.
 %the impulse response structure is similar, though the relative strength and directivity of the components may be different.
%For example, there may be greater emphasis on direct sound and ER rather than LR
In this section, we first give the details of our sound rendering algorithm (Section~\ref{sec:rendering}).
%In the remainder of this section, we describe the components of our spatial reverberation approach.
%First, we describe the details of the audio rendering algorithm .
Then, we describe how the sound rendering parameters can be robustly determined from a sound propagation impulse response with low sample rate (Section~\ref{sec:parameters}).

%%%%%%%%%%%%%%%%%%%%%%%%%%%%%%%%%%%%%%%%%%%%%%%%%%%%%%%%%%%%%%%%%%%%%%%%%%%%%%%%

\subsection{Sound Rendering} \label{sec:rendering}

% Reverb algorithm
	% Introduce prior work, e.g. Shroeder reverberator
	% Modifications:
		% Render in frequency bands
		% Render in SH domain

To render spatial reverberation, we extend the architecture proposed by Schroeder~\shortcite{schroeder1961}.
%but improve the quality of the results in several ways.
The Schroeder reverberator consists of $N_{comb}$ comb filters in parallel, followed by $N_{ap}$ all-pass filters in series.
We produce frequency-dependent reverberation by filtering the anechoic input audio, $s(t)$, into $N_{\bar{\omega}}$ discrete frequency bands using an all-pass Linkwitz-Riley $4$th-order crossover to yield a stream of audio for each band, $s_{\bar{\omega}}(t)$.
We use different feedback gain coefficients for each band in order to replicate the spectral content of the sound propagation IR and to produce different RT$_{60}$ at different frequencies.
To render directional reverberation, we extend the reverberator to operate in the spherical harmonic, rather than scalar, domain.
We render $N_{\bar{\omega}}$ frequency bands for each SH coefficient.
Therefore, the reverberation for each sound source consists of $(n + 1)^2 N_{\bar{\omega}}$ channels, where $n$ is the spherical harmonic order.

{\bf Input Spatialization:}
To model the directivity of the early reverberant impulse response,
we spatialize the input audio for each comb filter according to the directivity of the early IR.
We denote the spherical harmonic distribution of sound energy arriving at the listener for the $i$th comb filter as $X_{lm,i}$.
This distribution can be computed from the first few non-zero samples of the IR directivity, $X_{lm}(t)$, by interpolating the directivity at offset $t_{comb}^i$ past the first non-zero IR sample for each comb filter.
Given $X_{lm,i}$, we extract the dominant Cartesian direction from the distribution's 1st-order coefficients:
$\vec{x}_{max,i} = \text{normalize}(-X_{1,1,i}, -X_{1,-1,i}, X_{1,0,i})$~\cite{sloan2008}.
The input audio in SH domain for the $i$th comb filter is then given by evaluating the real SHs in the dominant direction and multiplying by the band-filtered source audio: $S_{\bar{\omega},lm}(t) = \frac{1}{N_{comb}} Y_{lm}(\vec{x}_{max,i}) s_{\bar{\omega}}(t)$.
We apply normalization factor $\frac{1}{N_{comb}}$ so that the reverberation loudness is independent of the number of comb filters.

{\bf SH Rotations:}
To simulate how sound tends to increasingly diffuse towards the end of the IR, we use SH rotation matrices in the comb filter feedback paths to scatter the sound.
The initial comb filter input audio is spatialized with the directivity of the early IR, and then the rotations progressively scatter the sound around the listener as the audio makes additional feedback loops through the filter.
At the initialization time, we generate a random rotation about the $x$, $y$, and $z$ axes for each comb filter and represent this rotation by $3 \times 3$ rotation matrix $\mathcal{R}_i$ for the $i$th comb filter.
The matrix is chosen  such that the rotation is in the range $[90^\circ, 270^\circ]$ in order to ensure there is sufficient diffusion.
Next, we build a SH rotation matrix, $J(\mathcal{R}_i)$, from $\mathcal{R}_i$ that rotates the SH coefficients of the reverberation audio samples during each pass through the comb filter.
The rotation matrix is computed according to the recurrence relations proposed by~\cite{ivanic1996}.
We can combine the rotation matrix with the frequency-dependent comb filter feedback gain $g_{comb,\bar{\omega}}^i$ to reduce the total number of operations required.
Therefore, during each pass through each comb filter, the delay buffer sample (a vector of $(n + 1)^2 N_{\bar{\omega}}$ values) is multiplied by matrix $J(\mathcal{R}_i) g_{comb,\bar{\omega}}^i$.
For the case of SH order $n = 1$, this operation is essentially a $4 \times 4$ matrix-vector multiply for each frequency band.
It may also be possible to use SH reflections instead of rotations to implement this diffusion process.

{\bf Directional Loudness:}
While the comb filter input spatializations model the initial directivity of the IR, and SH rotations can be used to model the increasing diffuse components in the later parts of the IR, we also need to model the overall directivity of the reverberation.
The weighted average directivity in SH domain for each frequency band, $\bar{X}_{\bar{\omega},lm}$, can be easily computed from the IR by weighting the directivity at each IR sample by the intensity of that sample:
\begin{equation}
\bar{X}_{\bar{\omega},lm} = \frac{1}{\int_0^\infty I_{\bar{\omega}}(t) dt} \int_0^\infty X_{\bar{\omega},lm}(t) I_{\bar{\omega}}(t) dt
\end{equation}
Given $\bar{X}_{\bar{\omega},lm}$, we need to determine a transformation matrix $D_{\bar{\omega}}$ of size $(n+1)^2 \times (n+1)^2$ that is applied to the $(n+1)^2$ reverb output SH coefficients in order to produce a similar directional distribution of sound for each frequency band $\bar{\omega}$.
This transformation can be computed efficiently using a technique for ambisonics directional loudness~\cite{kronlachner2014}.
The spherical distribution of sound $\bar{X}_{\bar{\omega},lm}$ is sampled for various directions in a spherical {\em t-design}, and then the discrete SH transform is applied to compute matrix $D_{\bar{\omega}}$.
$D_{\bar{\omega}}$ can then be applied to the SH coefficients of band $\bar{\omega}$ of each output audio sample after the last all-pass filter of the reverberator.

{\bf Early Reflections:}
The early reflections and direct sound are rendered in frequency bands using a separate delay interpolation module.
Each propagation path rendered in this manner produces $(n+1)^2 N_{\bar{\omega}}$ output channels that correspond to the SH basis function coefficients at $N_{\bar{\omega}}$ different frequency bands.
The amplitude for each channel is weighted according to the SH directivity for the path, where $x_{lm,j}$ are the SH coefficients for path $j$, as well as the path's pressure for each frequency band.
This enables our sound rendering architecture to handle area sound sources and diffuse reflections that are not localized in a single direction, as well as Doppler shifting for direct sound and early reflections.

%%%%%%%%%%%%%%%%%%%%%%%%%%%%%%%%%%%%%%%%%%%%%%%%%%%%%%%%%%%%%%%%%%%%%%%%%%%%%%%%

% Final spatialization
{\bf Spatialiation:}
Once the audio for all sound sources has been rendered in the SH domain and mixed together, it needs to be spatialized for the output audio format.
The audio for all sources in the SH domain is given by $q_{lm}(t)$.
After spatialization, the result is audio for each output channel is $q(t)$.
We propose two methods for spatialization: one using convolution with the listener's HRTF for binaural reproduction, and another using amplitude panning for surround-sound reproduction systems.

To spatialize the audio with the HRTF, the audio must be convolved with the listener's HRTF.
The HRTF, $H(\vec{x},t)$, is projected into the SH domain in a preprocessing step to produce SH coefficients $h_{lm}(t)$.
Since all audio is rendered in the world coordinate space, we need to apply the listener's head orientation to the HRTF coefficients before convolution to render the correct spatial audio.
If the current orientation of the listener's head is described by $3 \times 3$ rotation matrix $\mathcal{R}_L$, we construct a corresponding SH rotation matrix $\mathcal{J}(\mathcal{R}_L)$ that rotates HRTF coefficients from the listener's local to world orientation.
We then multiply the local HRTF coefficients by $\mathcal{J}$ to generate the world-space HRTF coefficients: $h_{lm}^{L}(t) = \mathcal{J}(\mathcal{R}_L) h_{lm}(t)$.
This operation is performed once for each simulation update.
The world-space reverberation, direct sound, and early reflection audio for all sources is then convolved with the rotated HRTF.
If the audio is rendered up to SH order $n$, the final convolution will consist of $(n+1)^2$ channels for each ear corresponding to the basis function coefficients.
After the convolution operation, the $(n+1)^2$ channels for each ear are summed to generate the final spatialized audio.
This operation is summarized in the following equation.
\begin{equation}
q(t) = \sum_{l=0}^n \sum_{m=-l}^l q_{lm}(t) \otimes \left[ \mathcal{J}(\mathcal{R}_L) h_{lm}(t) \right]
\end{equation}

It is also possible to efficiently spatialize the final audio using amplitude panning for surround-sound applications~\cite{pulkki1997}.
In that case, no convolution is required and our technique is even more efficient.
Starting with any ampliude panning model, e.g. vector-based amplitude panning (VBAP), we first convert the panning amplitude distribution for each speaker channel into the SH domain in a preprocessing step.
If the amplitude for a given speaker channel as a function of direction is represented by $A(\vec{x})$, we compute SH basis function coefficients $A_{lm}$ by evaluating the SH transform.
Like the HRTF, these coefficients must be rotated at runtime from listener-local to world orientation using matrix $\mathcal{J}(\mathcal{R}_L)$ each time the orientation is updated.
Then, rather than performing a convolution, we compute the dot product of the audio SH coefficients $q_{lm}(t)$ with the panning SH coefficients $A_{lm}$ for each audio sample:
\begin{equation}
q(t) = \sum_{l=0}^n \sum_{m=-l}^l q_{lm}(t) \left[ \mathcal{J}(\mathcal{R}_L) A_{lm} \right]
\end{equation}
With just a few multiply-add operations per sample, we can efficiently spatialize the audio for all sound sources using this method.

%%%%%%%%%%%%%%%%%%%%%%%%%%%%%%%%%%%%%%%%%%%%%%%%%%%%%%%%%%%%%%%%%%%%%%%%%%%%%%%%

\subsection{Reverberation Parameter Estimation} \label{sec:parameters}

In this section we describe how to acquire reverberation parameters that 
are needed to effectively render accurate reverberation. These parameters are computed using interactive ray tracing.
The input to our parameter estimation module is an impulse response (IR) generated by the sound propagation module that contains only the higher-order reflections (e.g. no early reflections or direct sound).
The IR consists of a histogram of sound intensity over time for various frequency bands, $I_{\bar{\omega}}(t)$, along with SH coefficients describing the spatial distribution of sound energy arriving at the listener position at each time sample, $X_{\bar{\omega},lm}(t)$.
The IR is computed at a low sample rate (e.g. $100$Hz) to reduce the noise in the Monte Carlo estimation of path tracing and to reduce memory requirements, since it is not necessary to use it for convolution at typical audio sampling rates (e.g. $44.1$ kHz).
This low sample rate is sufficient to capture the meso-scale structure of the IRs~\cite{kuttruff1993}.

{\bf Reverberation Time:}
The reverberation time, denoted RT$_{60}$, captures much of the sonic signature of an environment and corresponds to the time it takes for the sound intensity to decay by $60$dB from its initial amplitude.
We estimate the RT$_{60}$ from the intensity IR $I_{\bar{\omega}}(t)$ using standard techniques~\cite{iso3382}.
This operation is performed independently for each simulation frequency band to yield $RT_{60,\bar{\omega}}$.

Since the impulse response may contain significant amounts of noise, the RT$_{60}$ estimate may discontinuously change on each simulation update because the decay rate is sensitive to small perturbations.
To reduce the impact of this effect, we use temporal coherence similar to that proposed by~\cite{schissler2016a} to smooth the RT$_{60}$ over time with {\em exponential smoothing}.
Given a smoothing time constant $\tau$, we compute exponential smoothing factor $\alpha \in [0,1]$, then use $\alpha$ to filter the RT$_{60}$ estimate:
\begin{equation}
RT_{60,\bar{\omega}}^{n} = \mathbb{RT}_{60,\bar{\omega}}^{n} = \alpha \tilde{RT}_{60,\bar{\omega}}^{n} + (1-\alpha) \mathbb{RT}_{60,\bar{\omega}}^{n-1},
\end{equation}
where $RT_{60,\bar{\omega}}^{n}$ is the smoothed RT$_{60}$, $\tilde{RT}_{60,\bar{\omega}}^{n}$ is the RT$_{60}$ estimated from the current frame's IR,
$\mathbb{RT}_{60,\bar{\omega}}^{n-1}$ is the cached RT$_{60}$ value, and $\mathbb{RT}_{60,\bar{\omega}}^{n}$ is the cached value for the next frame.
By applying this smoothing, we reduce the variation in the RT$_{60}$ over time.
This also implies that the RT$_{60}$ may take about $\tau$ seconds to respond to an abrupt change in the scene.
However, since RT$_{60}$ is a global property of the environment and usually changes slowly, the perceptual impact of smoothing is less than that caused by noise in the RT$_{60}$ estimation.
Smoothing the RT$_{60}$ also makes our estimation more robust to noise in the IR caused by tracing only a few primary rays during sound propagation.

{\bf Direct to Reverberant Ratio:}
The direct to reverberant ratio (D/R ratio) determines how loud the reverberation should be in comparison to the direct sound.
The D/R ratio is important for producing accurate perception of the distance to sound sources in virtual environments~\cite{zahorik2002}.
%If the direct sound is much louder than the reverberation, the sound source seems to be very near to the listener, while if the reverberation is louder than the direct sound, the source seems to be farther away from the listener.
%Most previous work on using artificial reverberators for sound propagation has focused on determining the RT$_{60}$ parameter, but there is comparatively little work on accurately determining the D/R ratio in a robust manner.
The D/R ratio is described by the gain factor $g_{reverb}$ that is applied to the reverberation output such that the reverberation mixed with ER and direct sound closely matches the original sound propagation impulse response.

%We investigated several techniques for computing the reveberation loudness.
%The first used the $y$-intercept of the RT$_{60}$ least squares fit as the reverb gain.
%In some situations this method worked satisfactorily but in others (e.g. distant sound source) it was not very robust to noise in the impulse response.
%This would cause unnatural variation of at least $\pm 6$dB in the reverb loudness over time because the $y$-intercept is very sensitive to small changes in the RT$_{60}$ fitting.
%We also tried using the maximum of the impulse response as the reverb gain.
%Again, the maximum value in the IR varies significantly over time and produced rendering artifacts.
%This method also tended to underestimate the reverb loudness.

To robustly estimate the reverberation loudness from a noisy IR, we choose a method that has very little susceptibility to noise.
We found the most consistent metric to be the total intensity contained in the IR, i.e. $I_{\bar{\omega}}^{total} = \int_0^\infty I_{\bar{\omega}}(t) dt$.
To compute the correct reverberation gain, we derive a relationship between $I_{\bar{\omega}}^{total}$ and $g_{reverb}$.
This can be performed by finding the total intensity in the IR of a reverberator with $g_{reverb} = 1$, $I_{reverb}^{total}$.
Then, the gain factor of the reverberation output for each frequency band can be computed as the ratio of $I_{\bar{\omega}}^{total}$ to $I_{reverb,\bar{\omega}}^{total}$:
\begin{equation} \label{eq:reverb_gain}
g_{reverb,\bar{\omega}} = \sqrt{\frac{I_{\bar{\omega}}^{total}}{I_{reverb,\bar{\omega}}^{total}}}.
\end{equation}
The square root converts the ratio from intensity to the pressure domain.
To compute $I_{reverb,\bar{\omega}}^{total}$, given the RT$_{60}$, we model the reverberator's pressure envelope using a decaying exponential function $p_{reverb,\bar{\omega}}(t)$, derived from the definition of a comb filter:
\begin{equation} \label{eq:reverb_envelope}
p_{reverb,\bar{\omega}}(t) = \left\{
\begin{array}{lll}
      0 & :  t < 0, \\
      (g_{r,\bar{\omega}})^{t} & : t \ge 0,\\
\end{array}
\right.
\end{equation}
where $g_{r,\bar{\omega}}$ is the feedback gain for a comb filter with $t_{comb} = 1$ computed via the following equation:.
\begin{equation} \label{eq:comb_feedback_gain}
g_{comb}^i = 10^{-3 t_{comb}^i / {RT_{60}}}.
\end{equation}
We compute the total intensity of the reverberator by converting $p_{reverb}(t)$ to intensity domain by squaring, and then integrating from $0$ to $\infty$:
\begin{equation}
I_{reverb,\bar{\omega}}^{total} = \int_0^\infty (p_{reverb}(t))^2 dt = \frac{-1}{\ln{(g_{r,\bar{\omega}}^2)}} = \frac{RT_{60,\bar{\omega}}}{6\ln{10}}.
\end{equation}
Once $I_{reverb,\bar{\omega}}^{total}$ is computed, the gain factor for the reverberator can be computed using equation~\ref{eq:reverb_gain}.
Determining the reverberation loudness in this way is very robust to noise because it reuses as many Monte Carlo samples as possible from  ray tracing.

{\bf Reverberation Predelay:}
The reverberation predelay is the time in seconds that the first indirect sound arrival is delayed from $t=0$.
Usually, the predelay is correlated to the size of the environment.
%In larger environments, the first reflected sound is delayed more than in small environments because it must travel a greater distance from the source to the listener.
The predelay can be easily computed from the IR by finding the time delay of the first non zero sample, i.e. find $t_{predelay}$ such that $I_{\bar{\omega}}(t_{predelay}) \neq 0$ and $I_{\bar{\omega}}(t < t_{predelay}) = 0$ for all frequency bands.
This delay time is used as a parameter for the delay interpolation module of our sound rendering pipeline.
The input audio for the reverberator is read from the sound source's circular delay buffer at the time offset corresponding to the predelay.
This allows our approach to replicate the initial reverberation delay and give a plausible impression of the size of the virtual environment.

{\bf Reflection Density:}
%Another attribute of the impulse response is the density of reflections.
%In a highly reverberant indoor environment, the reflection density is high because the average length of a sound propagation path segment is small, so the reflections are spaced closer together in time.
%On the other hand, in a large open outdoor environment the distance between successive reflections is much larger and so the reflection density is less.
%Perceptually, high-density reflections result in a smooth sounding reverberation decay, while low-density reflections produce a coarse reverberation with distinct sound arrivals or echos.
To produce reverberation that closely corresponds to the environment, we need to model the reflection density, a parameter that is influenced by the size of the scene and controls whether the reverb is perceived as smooth decay or distinct echoes.
We perform this by gathering statistics about the rays traced during sound propagation, namely the {\em mean free path} of the environment.
The mean free path, $\bar{r}_{free}$, is the average unoccluded distance between two points in the environment and can be easily estimated during path tracing by computing the average distance that all rays travel.
Given $\bar{r}_{free}$, we can then choose reverberation parameters that produce echos every $\bar{r}_{free}/c$ seconds, where $c$ is the speed of sound.
To perform, we sample comb filter feedback delay times, $t_{comb}$, from a Gaussian distribution centered at $\bar{r}_{free}/c$ with standard deviation $\sigma = \frac{1}{3} \bar{r}_{free}/c$.
The feedback delay times are computed at the first initialization and  updated only when $\bar{r}_{free}/c$ deviates from the previous value by more than $2\sigma$ in order to reduce artifacts caused by resizing the delay buffers.

\section{Implementation} \label{sec:implementation}

{\bf Sound Propagation:}
We compute sound propagation in 4 logarithmically-spaced frequency bands: $0 - 176$Hz, $176 - 775$Hz, $775 - 3408$Hz, and $3408 - 22050$Hz.
To compute the direct sound, we use the Monte Carlo integration approach of~\cite{schissler2016b} to find the spherical harmonic projection of sound energy arriving at the listener.
The resulting SH coefficients can be used to spatialize the direct sound for area sound sources using our rendering approach.
To compute early reflections and late reverberation, we use backward path tracing from the listener because it scales well with the number of sources~\cite{schissler2016c}.
Forward or bidirectional ray tracing may also be used~\cite{cao2016}.
We augment the path tracing using {\em diffuse rain}, a form of next-event estimation, in order to improve the path tracing convergence~\cite{schroder2007}.
To handle early reflections, we use the first $2$ orders of reflections in combination with the {\em diffuse path cache} temporal coherence approach to improve the quality of the early reflections when a small number of rays are traced~\cite{schissler2014}.
We improve on the original cache implementation by augmenting it with spherical-harmonic directivity information for each path.
For reflections over order $2$, we accumulate the ray contributions to an {\em impulse response cache}~\cite{schissler2016a} that utilizes temporal coherence in the late IR.
The computed IR has a low sampling rate of $100$Hz that is sufficient to capture the meso-scale IR structure.
We use this IR to estimate reverberation parameters.
Due to the low IR sampling rate, we can trace far fewer rays to maintain good sound quality.
We emit just $50$ primary rays from the listener on each frame and propagate those rays to reflection order $200$.
If a ray escapes the scene before it reflects $200$ times, the unused ray budget is used to trace additional primary rays.
Therefore, our approach may emit more than $50$ primary rays on outdoor scenes, but always traces the same number of ray path segments.
The two temporal coherence data structures (for ER and LR) use different smoothing time constants $\tau_{ER} = 1$s and $\tau_{LR} = 3$s, in order to reduce the perceptual impact of lag during dynamic scene changes.
Our system does not currently handle diffraction effects, but it would be relatively simple to augment the path tracing module with a probabalistic diffraction approach~\cite{stephenson2010}, though with some extra computational cost.
Other diffraction algorithms such as UTD and BTM require significantly more computation and would not be as suitable for low-cost sound propagation.
Sound propagation is computed using 4 threads on a 4-core desktop machine, or 2 threads on the Google Pixel XL\texttrademark mobile device.

{\bf Auralization:}
Auralization is performed using the same frequency bands that are used for sound propagation.
We make extensive use of SIMD vector instructions to implement rendering in frequency bands efficiently: bands are interleaved and processed together in parallel.
The audio for each sound source is filtered into those bands using a time-domain Linkwitz-Riley $4$th-order crossover and written to a circular delay buffer.
The circular delay buffer is used as the source of prefiltered audio for direct sound, early reflections, and reverberation.
The direct sound and early reflections read delay taps from the buffer at delayed offsets relative to the current write position.
The reverberator reads its input audio as a separate tap with delay $t_{predelay}$.
In the reverberator, we use $N_{comb} = 8$ comb filters and $N_{ap} = 4$ all-pass filters.
This improves the subjective quality of the reverberation versus the orignal Schroeder design~\cite{schroeder1961}.

We use different a spherical harmonic order for the different sound propagation components.
For direct sound, we use SH order $n = 3$ because the direct sound is highly directional and perceptually important.
For early reflections, we use SH order $n = 2$ because the ER are slightly more diffuse than direct sound and so a lower SH order is not noticeable.
For reverberation, we use SH order $n = 1$ because the reverb is even more diffuse and less important for localization.
When the audio for all components is summed together, the unused higher-order SH coefficients are assumed to be zero.
This configuration provided the best tradeoff between auralization performance and subjective sound quality by using higher-order spherical harmonics only where needed.
%An analogous approach has been used to adaptively determine the SH order for spatial IR construction~\cite{schissler2017b}.

To avoid rendering too many early reflection paths, we apply a sorting and prioritization step to the raw list of the paths.
First, we discard any paths that have intensity below the listener's threshold of hearing.
Then, we sort the paths in decreasing intensity order and use only the first $N_{ER} = 100$ among all sources for audio rendering.
The unused paths are added to the late reverberation IR before it is analyzed for reverb parameters.
This limits the overhead for rendering early reflections by rendering only the most important paths.
Auralization is implemented on a separate thread from the sound propagation and therefore is computed in parallel.
The auralization state is synchronously updated each time a new sound propagation IR is computed.
\section{Results \& Analysis} \label{sec:results}

\begin{table*}
\centering
\small
    \begin{tabular}{lcc|cccc|cc}
	\toprule
  			      & \multicolumn{2}{c}{\bf Scene Complexity} & \multicolumn{4}{c}{\bf Propagation} & \multicolumn{2}{c}{\bf Auralization} \\
    Scene	      & \#Tris     & \#Sources   & Ray Tracing (ms) & IR Analysis (ms)  & Total (ms)  & Speedup      & \% Max     & Speedup        \\
    \midrule
    City          & 1,001,860  & 13          & 6.85        & 0.23         & 7.08   & {\bf 9.3}    & 11.6 & {\bf 1.9}    \\
    Hospital      & 64,786     & 12          & 7.18        & 0.18         & 7.35   & {\bf 9.2}    & 10.7 & {\bf 1.7}    \\
    Space Station & 49,258     & 23          & 12.75       & 0.45         & 13.20  & {\bf 10.2}   & 20.2 & {\bf 1.6}    \\
    Sub Bay       & 402,477    & 19          & 14.10       & 0.50         & 14.60  & {\bf 12.8}   & 18.7 & {\bf 3.1}    \\
    Tuscany       & 371,157    & 14          & 10.82       & 0.21         & 11.03  & {\bf 9.3}    & 12.5 & {\bf 1.8}    \\
    \midrule
    Hospital (mobile)      & 64,786     & 8          & 83.71       & 0.17      & 83.89      & {\bf 15.5}      & 19.4 & {\bf 3.9}    \\
    Space Station (mobile) & 49,258     & 7          & 66.32       & 0.10      & 66.42      & {\bf 12.1}      & 18.7 & {\bf 2.0}    \\
    \bottomrule
    \end{tabular}
\caption{The main results of our sound propagation and auralization approach.
In the upper part of the table, we show the performance results using 4 ray tracing threads and 1 auralization thread on a desktop Intel i7 4770k CPU.
In the lower part, we show the results for benchmarks on a Google Pixel XL\texttrademark mobile device with 2 ray tracing threads and 1 auralization thread.
Our approach is able to achieve a significant speedup of about 10x over convolution-based rendering on desktop CPUs, and is the first to demonstrate interactive dynamic sound propagation on a low-power mobile CPU.
}\label{table:main_results}
\vspace*{-0.15in}
\end{table*}

We evaluated our approach on a desktop machine using five benchmark scenes that are summarized in Table~\ref{table:main_results}.
These scenes contain between 12 and 23 sound sources and have up to 1 million triangles as well as dynamic rigid objects.
For two of the five scenes, we also prepared versions with less sound sources that were suitable for running on a mobile device.
In Table~\ref{table:main_results}, we show the main results of our technique, including the time taken for ray tracing, analysis of the IR (determination of reverberation parameters), as well as auralization.
The auralization time is reported as the percentage of real time needed to render an equivalent length of audio, where $100\%$ indicates the rendering thread is fully saturated.
The results for the five large scenes were measured on a 4-core Intel i7 4770k CPU, while the results for the mobile scenes were measured on a Google Pixel XL\texttrademark phone with a 2+2 core Snapdragon 821 chipset.

\begin{figure}
\centering
\includegraphics[width=0.96\columnwidth]{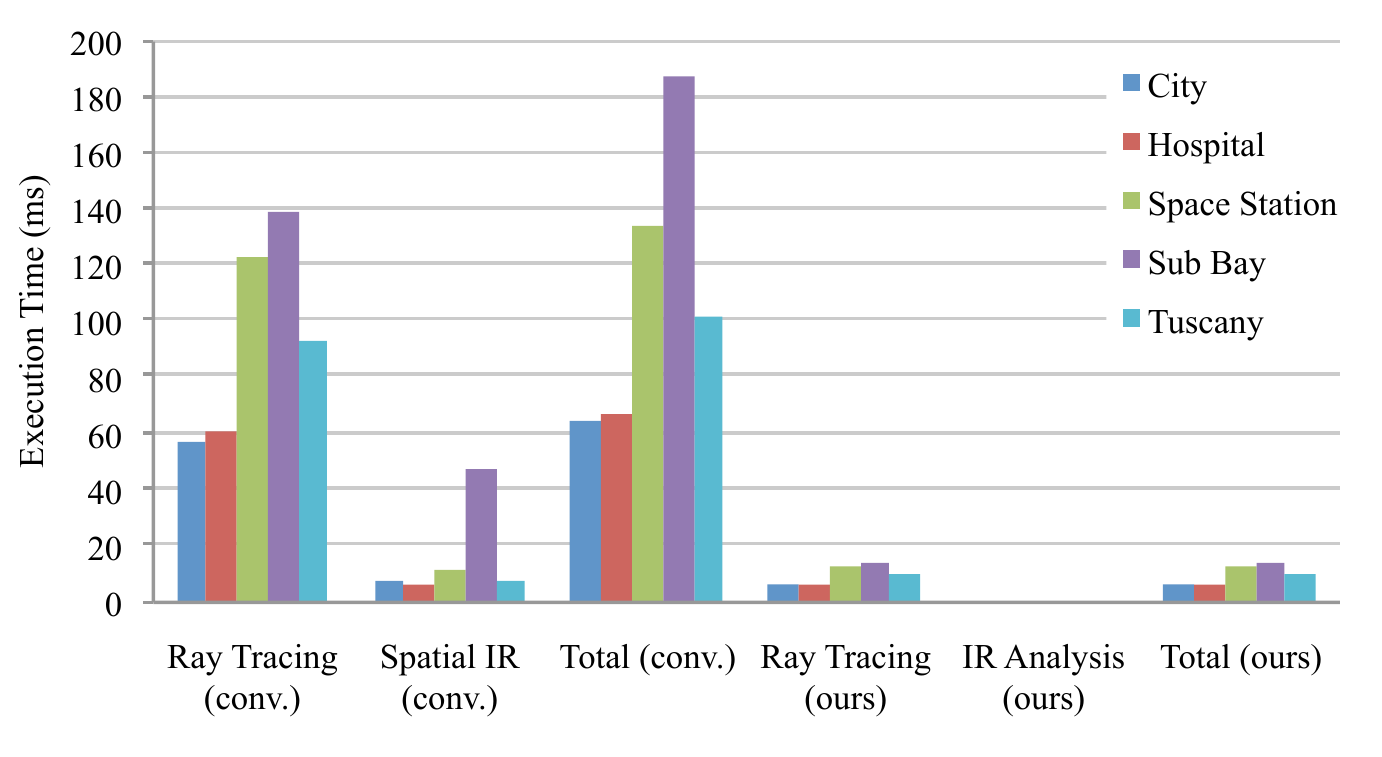}
\caption{A comparison between the sound propagation performance of our approach and a traditional convolution-based architecture on a desktop machine.
We report the time taken on various scenes for the components of sound propagation: ray tracing and spatial IR construction/analysis.}\label{fig:propagation_vs_time}
\end{figure}

{\bf Performance:}
The sound propagation performance is reported in Table~\ref{table:main_results}.
On the desktop machine, roughly $6 - 14$ms is spent on ray tracing in the five main scenes.
This corresponds to about $0.5 - 0.75$ms per sound source.
The ray tracing performance scales linearly with the number of sound sources and is typically a logarithmic function of the geometric complexity of the scene.
On the mobile device, ray tracing is substantially slower, requiring about $10$ms for each sound source.
This may be because our ray tracer is more optimized for Intel than ARM CPUs.
We also report the time taken to analyze the impulse response and determine reverberation parameters.
On both the desktop and mobile device, this component takes about $0.1 - 0.5$ms.
The total time to update the sound rendering system is $7 - 14$ms on the desktop and $66-84$ms on the mobile device.
As a result, the latency of our approach is low enough for interactive applications and is the first to enable dynamic sound propagation on a low-power mobile device.

In comparison, the performance of traditional convolution-based rendering is substantially slower.
Figure~\ref{fig:propagation_vs_time} shows a comparison between the sound propagation performance of state of the art convolution-based rendering and our approach.
Convolution-based rendering requires about $500$ rays to acheive sufficient sound quality without unnatural sampling noise when temporal coherence is used~\cite{schissler2016a}.
On the other hand, our approach is able to use only $50$ rays due to its robust reverb parameter estimation and rendering algorithm.
This provides a substaintial speedup of $9.2 - 12.8$x on the desktop machine, and a $12.1 - 15.5$ speedup on the mobile device.
A significant bottleneck for convolution-based rendering is the computation of spatial impulse responses from the ray tracing output, which requires time proportional to the IR length.
The Sub Bay scene has the longest impulse response and has a spatial IR cost of $48$ms that is several times that of the other scenes.
However, our approach requires less than a millisecond to analyze the IR and update the reverberation parameters.

\begin{figure}
\centering
\includegraphics[width=0.96\columnwidth]{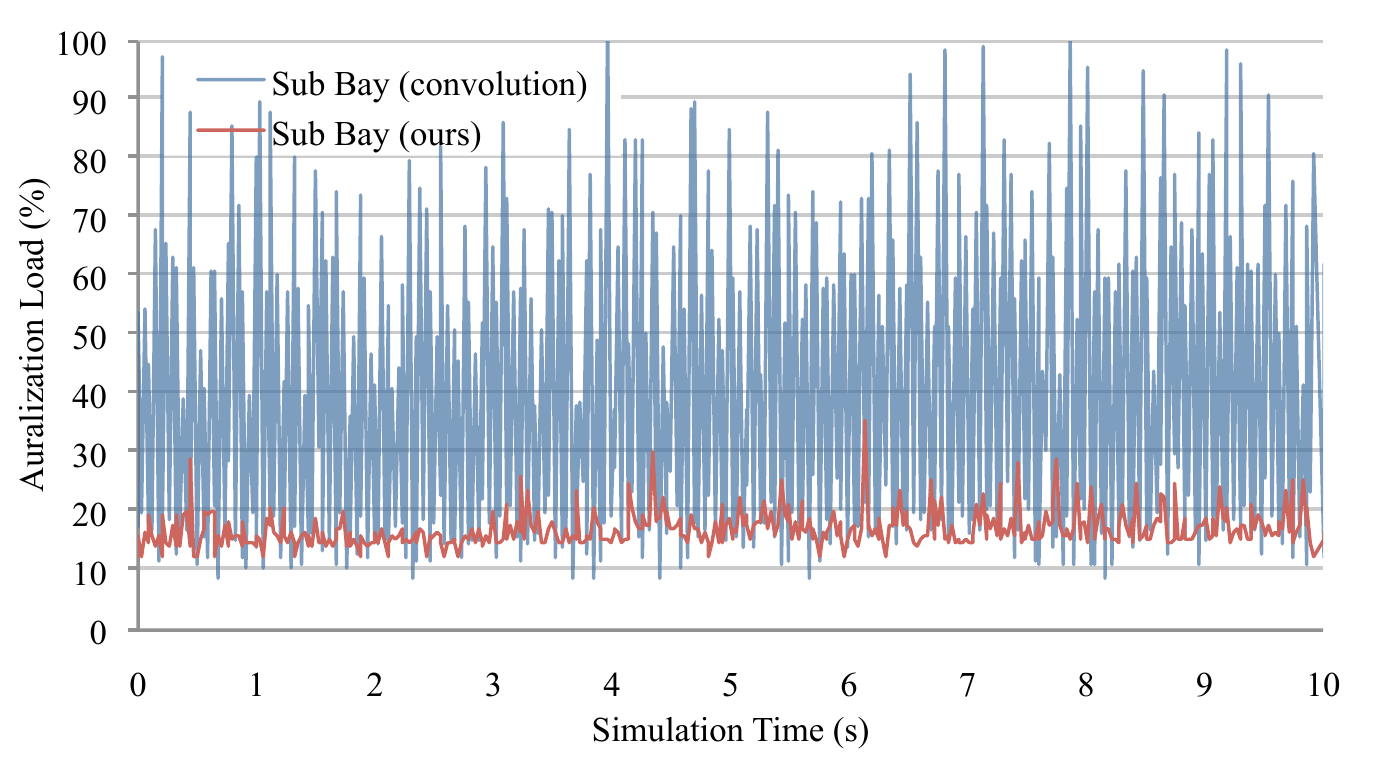}
\caption{A performance comparison between our reverberation rendering algorithm and a traditional convolution-based rendering architecture on a single thread.
The average cost of our approach per source is less than half that of convolution and the performance is much more consistent over time.
This makes clicks and pops less likely to happen with our approach versus convolution when rendering many sources, since the maximum rendering time is reduced.}\label{fig:render_vs_time}
\vspace*{-0.15in}
\end{figure}

\begin{figure}
\centering
\includegraphics[width=0.96\columnwidth]{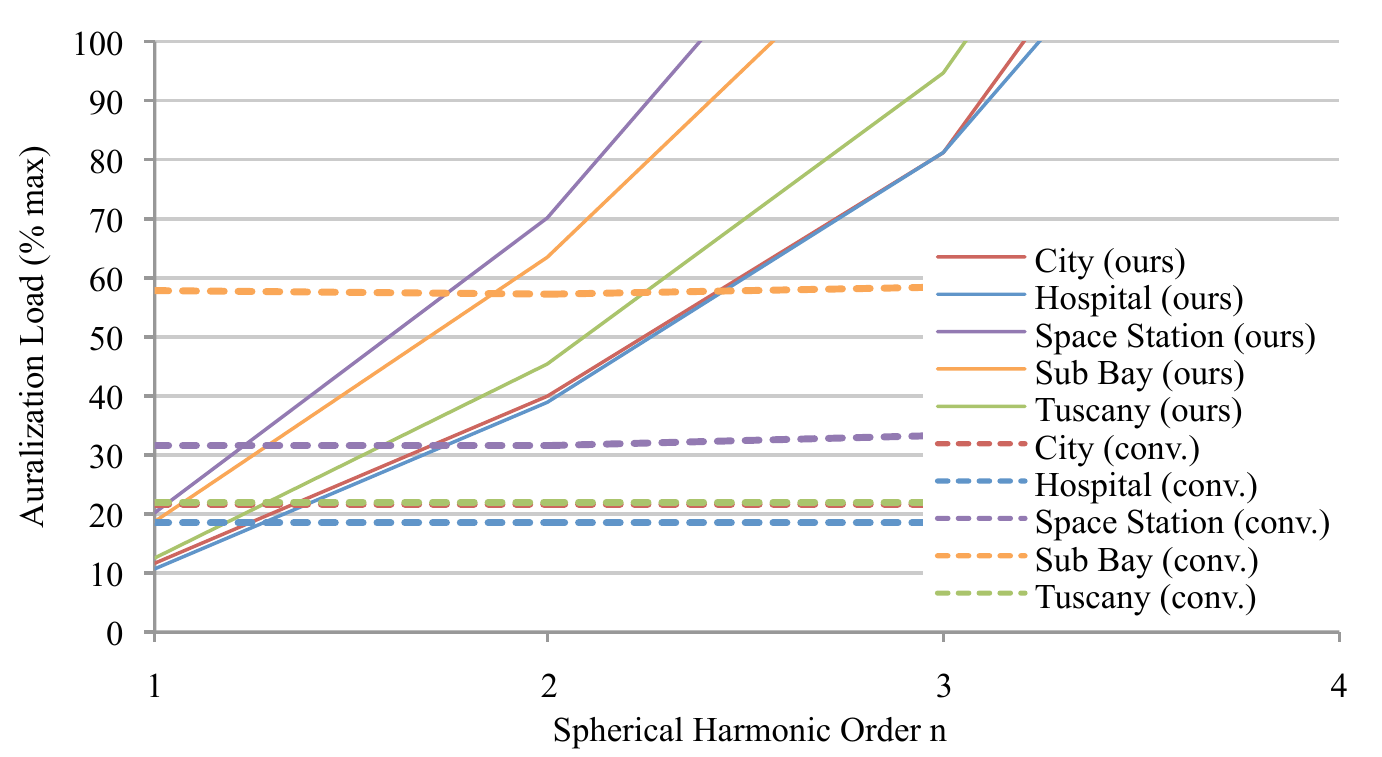}
\caption{The rendering performance of our reverberation algorithm varies based on the spherical harmonic order used.
We observe quadratic scaling with respect to the SH order.
For SH order $n = 1$, our approach is about 2x faster than a convolution-based renderer.}\label{fig:render_vs_sh}
\vspace*{-0.15in}
\end{figure}

With respect to the auralization performance, our approach uses $11 - 20$\% of one thread to render the audio.
In comparison, an optimized low-latency convolution system requires about $1.6 - 3.1$x more computation.
A significant drawback of convolution is that the computational load is not constant over time, as shown in Figure~\ref{fig:render_vs_time}.
Convolution has a much higher maximum computation than our auralization approach and therefore is much more likely to produce audio artifacts due to not meeting real-time requirements.
A traditional convolution-based pipeline also requires convolution channels in proportion to the number of sound sources.
As a result, convolution becomes impractical for more than a few dozen sound sources~\cite{lentz2007}.
On the other hand, our approach uses only a constant number of convolutions per listener for spatialization with the HRTF, where the number of convolutions is $2(n+1)^2$.
This means that for SH order $n=3$, we render only 32 channels of convolution with a very short HRTF impulse response, whereas a convolution-based system would have to convolve with an impulse response over 100x longer for each sound source and channel.
If not using HRTFs, our approach requires no convolutions.
The performance of our auralization algorithm is strongly dependent on the spherical harmonic order.
In Figure~\ref{fig:render_vs_sh}, we demonstrate quadratic scaling for SH orders $1-4$.
Our approach is faster than convolution-based rendering for $n=1$, but becomes impractical at higher SH orders.
However, reverberation is smoothly directional, so low order spherical harmonics are sufficient to capture most directional effects.

{\bf Memory:}
A further advantage of our technique is that the memory required for impulse responses and convolution is greatly reduced.
We store the IR at $100$Hz sample rate, rather than $44.1$kHz.
This provides a memory savings of about $441$x for the impulse responses.
Our approach also omits convolution with long impulse responses, which requires at least 3 IR copies for low-latency interpolation.
Therefore, our approach uses significant memory for only the delay buffers and reverberator, totaling about $1.6$MB per sound source.
This is a total memory reduction of about $10$x versus a traditional convolution-based renderer.

\begin{figure}
\centering
\includegraphics[width=0.96\columnwidth]{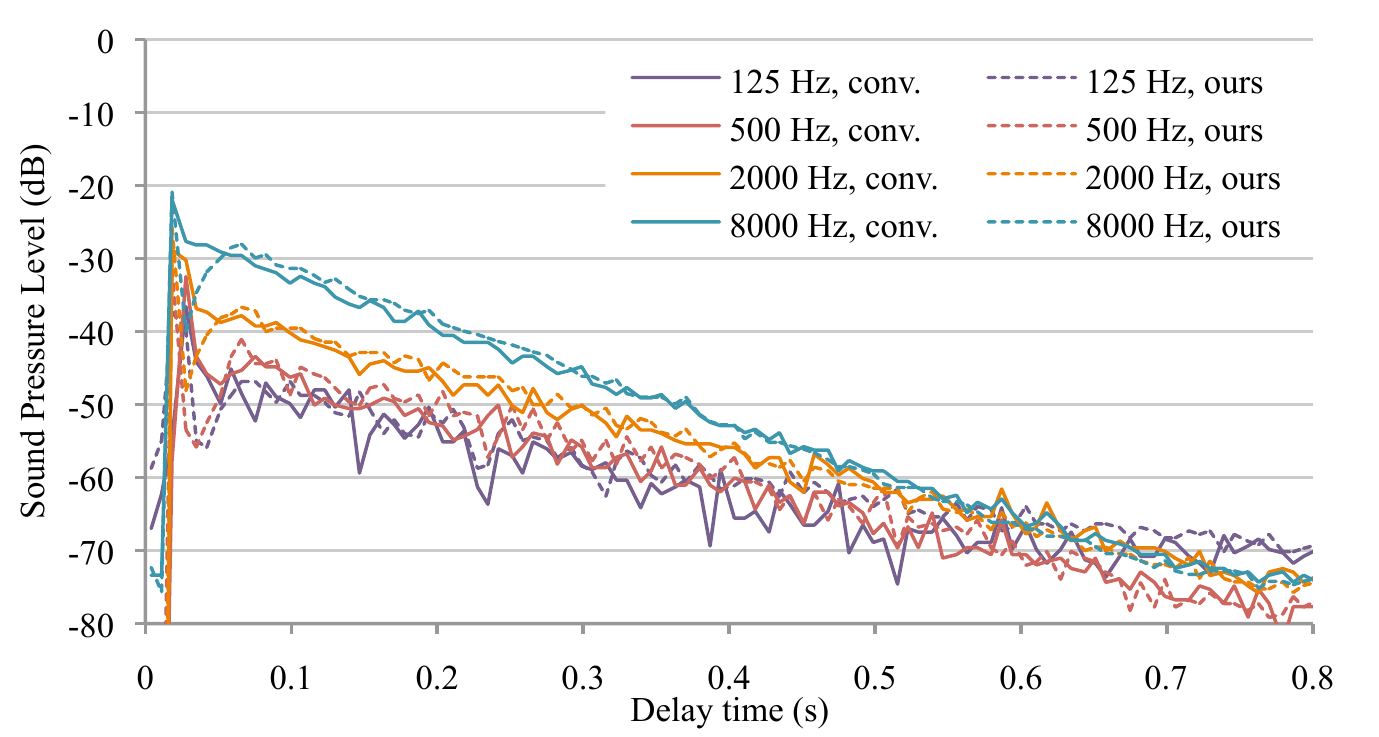}
\caption{A comparison between an impulse response generated by our spatial reverberation approach and a high-quality impulse response computed using previous approaches for frequency bands centered at $125$Hz, $500$Hz, $2000$Hz, and $8000$Hz.
Our approach closely matches the overall shape and energy distribution of the actual IR, with an average error of $3.4$dB, $2.7$dB, $1.6$dB, and $1.2$dB for the respective frequency bands.}\label{fig:ir_comparison}
\end{figure}

{\bf Validation:}
In Figure~\ref{fig:ir_comparison} we compare the impulse response generated by our method to the impulse response generated by a convolution-based sound rendering system in the space station scene.
We show the envelopes of the pressure impulse response for 4 frequency bands, which were computed by applying the Hilbert transform to the band-filtered IRs.
Our approach closely matches the overall shape and decay rate of the convolution impulse response at different frequencies, and preserves the relative levels between the frequencies.
In addition, our approach generates direct sound that corresponds to the convolution IR.
The average error between the IRs is between $1.2$dB and $3.4$dB across the frequency bands, with the error generally increasing at lower frequencies where there is more noise in the IR envelopes.
With respect to standard acoustic metrics like RT$_{60}$, $C_{80}$, $D_{50}$, $G$, and $TS$, our method is very close to the convolution-based method.
For RT$_{60}$, the error is in the range of $5-10$\%, which is close to the just noticeable difference of $5$\%.
For $C_{80}$, a measure of direct to reverberant sound, the error between our method and convolution-based rendering is $0.6 - 1.3$dB.
The error for $D_{50}$ is just $2-10$\%, while $G$ is within $0.2 - 0.8$dB.
The center time, $TS$, is off by just $1-7$ms.
Overall, our method generates audio that closely matches convolution-based rendering on a variety of comparison metrics.

\section{User Evaluation} \label{sec:study}

We also evaluated the subjective quality of our approach with a simple online user study.
In the study, we compared three methods for rendering the sound: {\em conv}, a convolution-based approach; {\em reverb}, an artificial reverberator with diffuse directivity rendered in 2-channel stereo format; and {\em our} approach.
The study investigated the subjective differences between the pairings {\em conv} and {\em our}, {\em reverb} and {\em our}, as well as the reverse pairings for the five benchmark scenes.
Therefore, the study consisted of a total of 20 paired comparisons.
The participants were shown short videos for the comparisons where the virtual listener was spawned at a static location receiving only indirect sound from occluded sound sources.
In every video, the listener rotates their head from side to side to demonstrate the spatial audio.
After watching each video pairing, the user answered a short questionaire comparing the videos.
The questions were:
\begin{itemize}
\vspace{-0.1em}
\item In which video could you better localize the sound source?
\item In which video was the sound more spacious?
\item In which video was the sound more realistic?
\item Which video did you prefer?
\end{itemize}
Each question was answered on a Likert scale from 1 to 11, where 1 indicates a strong preference for the left video, 11 indicates a strong preference for the right video, and 6 indicates no preference.

\begin{figure}[t!]
    \centering
    \begin{subfigure}[b]{\columnwidth}
        \centering
        \caption{{\em conv} vs. {\em our} approach}
        \includegraphics[width=\columnwidth]{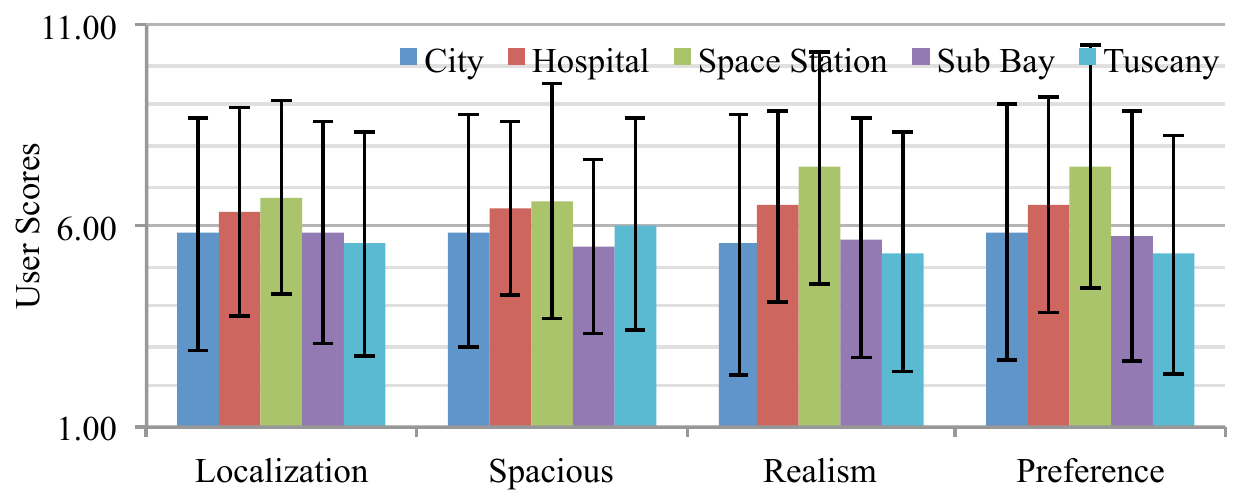}
    \end{subfigure}
    \begin{subfigure}[b]{\columnwidth}
        \centering
        \caption{{\em reverb} vs. {\em our} approach}
        \includegraphics[width=\columnwidth]{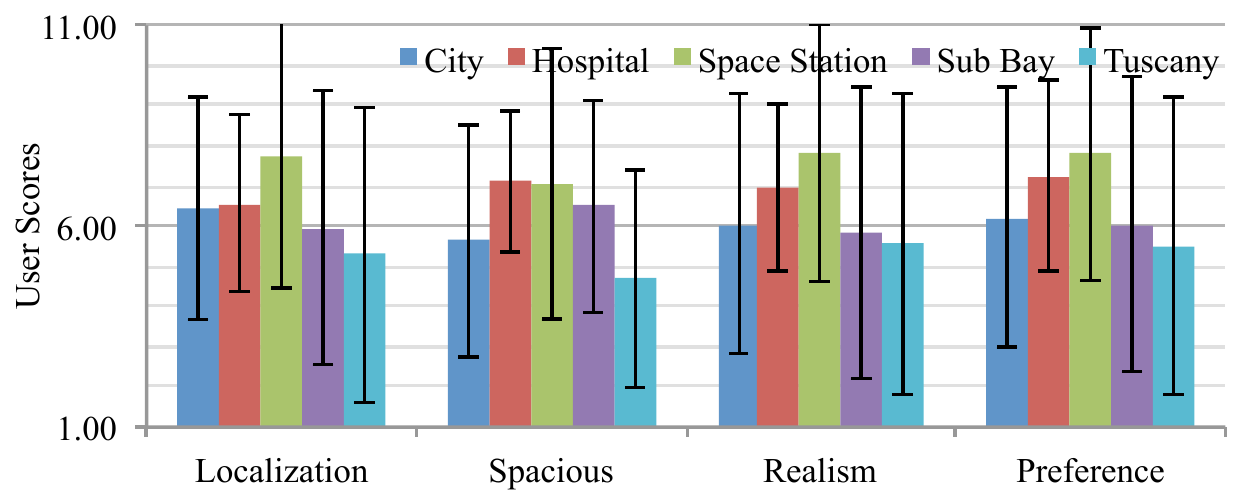}
    \end{subfigure}
    \caption{The main results of the user evaluation of our method. A score of 1 indicates a strong preference for the first method ({\em conv} or {\em reverb}), while a score of 11 indicates a strong preference for the {\em our} method.
    A score of 6 indicates no preference.
    Error bars indicate the standard deviation of the responses.
    Overall, the results are inconclusive.
    } \label{fig:study_results}
\end{figure}

{\bf User Evaluation Results:}
There were $20$ subjects who completed the study.
The main results for the two comparisons are presented in Figure~\ref{fig:study_results}.
Overall, there is no significant preference for any rendering method on any question.
This suggests either that the subjects have difficulty discerning between the rendering approaches or that the audio for all methods is very similar.
There is a consistent small preference for {\em our} approach on the space station scene, indicating that the type of scene influences user preferences and that some scenes may have more obvious differences.
The results of the study suggest that the average listener does not have the ability to hear subtle differences in the sound.
In the future, we would like to perform additional listening tests with expert listeners (e.g. individuals skilled at audio mixing) in order to improve the quality of the evaluation.

\section{Conclusions and Limitations} \label{sec:conclusion}

In this work we have presented a novel sound propagation and rendering architecture based on spatial artificial reverberation.
Our approach uses a spherical harmonic representation to efficiently render directional reverberation, and robustly estimates the reverberation parameters from a coarsly-sampled impulse response.
The result is that our method can generate plausible sound that closely matches the audio produced using more expensive convolution-based techniques, including directional effects.
In practice, our approach can generate plausible sound that closely matches the audio generated by state of the art methods based on convolution-based sound rendering pipeline.
We have evaluated its performance on complex scenarios and observe more than an order of magnitude speedup over convolution-based rendering.
We believe ours is the first approach that can generate rendering interactive dynamic physically-based sound on current mobile devices.

%The result is that our method can generate plausible sound that closely matches the audio produced using more expensive convolution-based techniques, including directional effects.
%We have evaluated the performance on a variety of scenes and observe an improvement of $9.2 - 15.5$x for sound propagation and $1.6 - 3.9$x for sound rendering.
%We also conducted a user evaluation to study the subjective differences between our approach and convolution-based sound rendering.
%The results suggest that the differences between our approach and the previous one are insignificant.
%Due to the significant performance improvements of our technique, 

Our approach has some limitations.
Because we use artificial reverberation, rather than convolution to render the sound, our approach is not capable of reproducing the fine-scale detail in a $44.1$kHz IR.
All the other limitations of geometric acoustics (e.g. inaccuracies at lower frequencies) can also arise in our approach. 
Due to the use of low-order SHs, our approach may not be able to represent very sharp directivities, though the SH order can be increased.
%with a greater computational cost.
Recent work has shown that $4$th order is sufficient for accurate localization of point sources with individualized HRTFs~\cite{romigh2015}, and there are no significant differences between $1$st, $2$nd, and $3$rd-order ambisonics in the reproduction of room acoustics.
We make extensive use of temporal coherence techniques to improve the quality of the sound by trading some interactivity, and this can result in a slow response if there are fast dynamic changes.
%Since we use fewer rays to compute a low pass IR, it can result in additional errors in UTD based diffraction algorithms.
%As a result, our technique may take a few seconds to react if there is a fast dynamic change in the scene, such as a large door opening or closing or a fast-moving sound source.
%However, in most cases the negative perceptual impact of temporal smoothing is small compared to the noise artifacts it prevents.
%In extreme scene changes, the temporal coherence data strutures can also be reset to avoid lag.

There are many avenues for future work.
One possibility is to perform a more detailed perceptual evaluation and study the impact that the fine structure of the IR has on the subjective quality of the sound.
We would also like to investigate other reverberator designs that may be more efficient to render or may improve the sound quality.
We used Schroeder-type reverberator which produces a constant echo density, while most real rooms have an exponentially increasing echo density~\cite{schroeder1961}.
A different design may produce equal or better reverberation quality with less computational resources.
We would also like to investigate efficient ways to incorporate diffraction into our sound propagation framework without significantly impacting the performance.
%One possibility is to use a probabalistic diffraction technique that can be efficiently integrated into the path tracer~\cite{stephenson2010}.
%We would also like to investigate if bidirectional ray tracing provides any improvements in quality over ray tracing from the listener, especially because fewer rays may be required to produce a noise-free IR.
% \input{appendix.tex}

%\section*{Acknowledgements}
%We would like to thank the anonymous reviewers for comments and feedback.
%We would also like to thank Valve Corporation for permission to use Half-Life 2 sound effects in the Office and Refinery scenes.
%This research was supported in part by
%	the Link Foundation Fellowship in Advanced Simulation and Training,
%	ARO Contracts W911NF-10-1-0506, W911NF-12-1-0430, W911NF-13-C-0037,
%	and the National Science Foundation (NSF awards 0917040, 1320644),

%\begin{small}
%\bibliographystyle{acmsiggraph}
\bibliographystyle{ACM-Reference-Format}
\bibliography{paper}
%\end{small}

%\clearpage
%\newpage
%\pagenumbering{arabic}
%\input{a-appendix}
\newpage

\section*{Supplemental Material}

%%%%%%%%%%%%%%%%%%%%%%%%%%%%%%%%%%%%%%%%%%%%%%%%%%%%%%%%%%%%%%%%%%%%%%%%%%%%%%%%

In this supplemental document, we present additional background that includes:
\begin{itemize}
\item Details on previous artificial reverberation algorithms.
\item Description of the major components of the impulse response.
\end{itemize}

%%%%%%%%%%%%%%%%%%%%%%%%%%%%%%%%%%%%%%%%%%%%%%%%%%%%%%%%%%%%%%%%%%%%%%%%%%%%%%%%

\begin{figure}
\centering
\includegraphics[width=0.96\columnwidth]{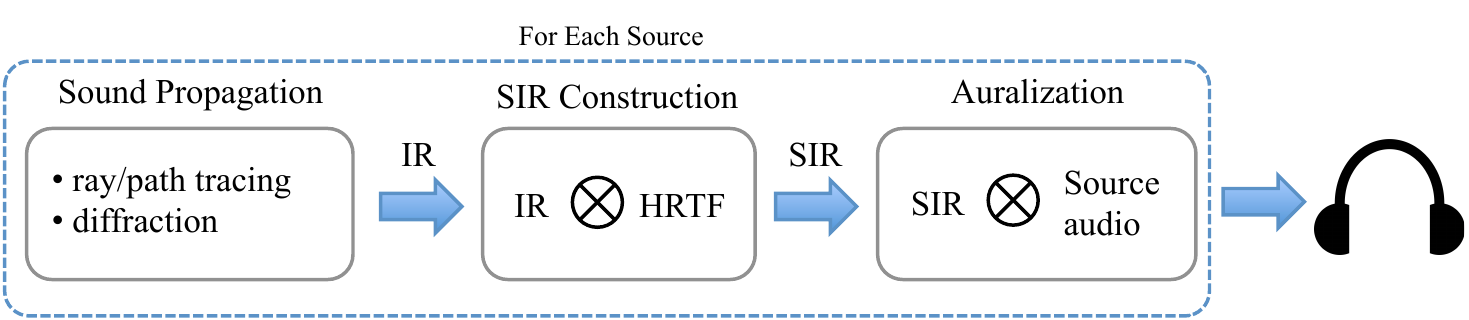}
\caption{A traditional sound propagation rendering pipeline that uses convolution with the impulse response.
The sound propagation module computes an impulse response (IR) based on the source and listener positions within the scene.
Then, the IR is converted into a spatial impulse response (SIR) by convolution with the listener's head related transfer function (HRTF).
Finally, the SIR is convolved with the anechoic source audio to generate the output sound.}\label{fig:pipeline_old}
\end{figure}

%%%%%%%%%%%%%%%%%%%%%%%%%%%%%%%%%%%%%%%%%%%%%%%%%%%%%%%%%%%%%%%%%%%%%%%%%%%%%%%%

\section{Artificial Reverberation Algorithms} \label{sec:reverb}

\begin{figure}
\centering
\includegraphics[width=0.96\columnwidth]{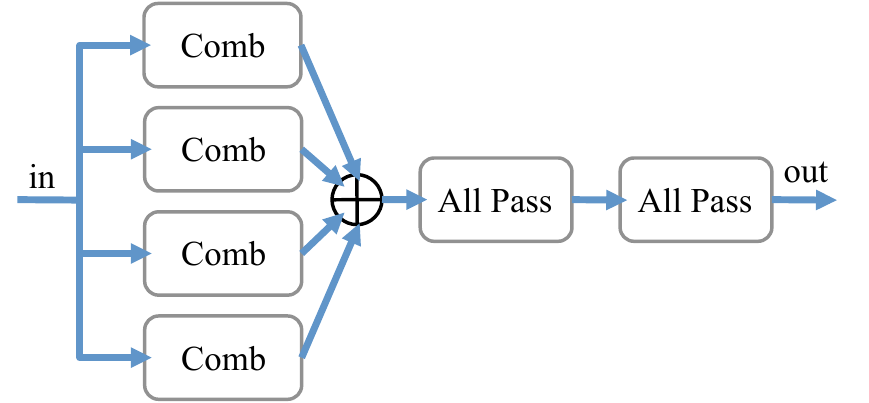}
\caption{A Schroeder-type reverberator consisting of $N_{comb}$ comb filters in parallel, followed by $N_{ap}$ all-pass filters in series. We show the original design proposed by Schroeder, with $N_{comb}=4$ and $N_{ap}=2$.
In this design, comb filters produce the shape of the reverberation decay, while the all-pass filters increase the number of echoes and help to produce a smooth reverberation decay.
}\label{fig:schroeder_reverb}
\end{figure}

\begin{figure}
\centering
\includegraphics[width=0.49\columnwidth]{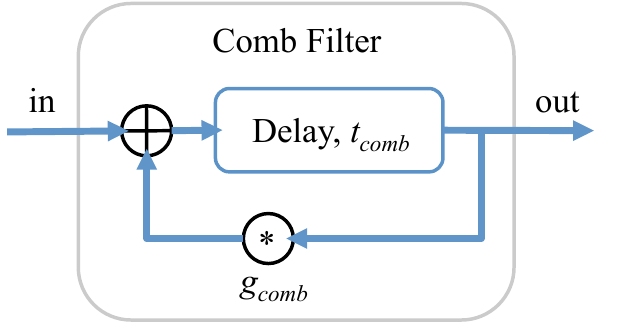}
\includegraphics[width=0.49\columnwidth]{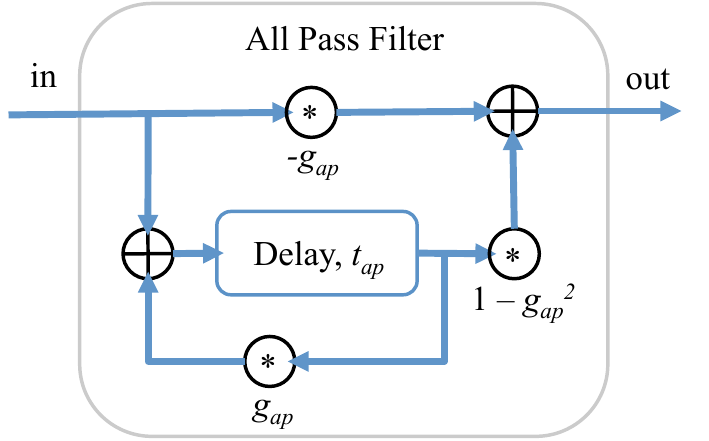}
\caption{The block diagrams for infinite impulse response filters commonly used in rendering artificial reverberation.
Comb filters produce a series of exponentially-decaying echoes with a period of $t_{comb}$, where each echo has magnitude of $g_{comb}$ times the previous echo.
All-pass filters produce a similar series of echoes with period $t_{ap}$, but with a flat magnitude response.
The feedback gain parameters $g_{comb}$ and $g_{ap}$ must be in the range $[0,1)$ for the filters to be stable.
}\label{fig:filters}
\end{figure}

Various previous approaches have been proposed for rendering artificial reverberation.
One of the earliest and most widely used reverberator designs was proposed by Schroeder~\cite{schroeder1961}.
This reverberator, shown in Figure~\ref{fig:schroeder_reverb}, consists of $N_{comb}$ comb filters in parallel, followed by $N_{ap}$ all-pass filters in series.
Schematic diagrams of comb and all-pass filters are shown in Figure~\ref{fig:filters}.
The $i$th comb filter produces a series of exponentially decaying echoes with a period of $t_{comb}^i$ by passing the input audio through a circular delay buffer of length $t_{comb}^i$.
During audio rendering, samples are read from the delay buffer and multiplied by feedback gain factor $g_{comb}^i \in [0,1)$.
The result is mixed with the current input audio and then written back into the delay buffer at the same location.
Therefore, each comb filter generates an infinite impulse response that has a sequence of decaying echoes where each echo is a factor of $g_{comb}$ less than the previous echo.
Usually, the values of $t_{comb}^i$ are chosen so that no two $t_{comb}^i$ are close to an integer multiple of each other in order to reduce metallic ringing artifacts.
The values of $g_{comb}^i$ are computed so that the comb filters have a decay rate that is consistent with the room's $RT_{60}$:
\begin{equation} \label{eq:comb_feedback_gain}
g_{comb}^i = 10^{-3 t_{comb}^i / {RT_{60}}}
\end{equation}
While comb filters generate the overall envelope of the impulse response, they do not generate a sufficient number of echoes to produce smooth reverberation.
To improve the results, all pass filters take the output of the comb filters and generate many small additional echoes with a shorter period.
The all-pass filters have similar parameters, $t_{ap}^i$ and $g_{ap}^i$, and are used to further increase the echo density.

%%%%%%%%%%%%%%%%%%%%%%%%%%%%%%%%%%%%%%%%%%%%%%%%%%%%%%%%%%%%%%%%%%%%%%%%%%%%%%%%

\section{Impulse Response Structure} \label{sec:ir_structure}

\begin{figure}
\centering
\includegraphics[width=0.96\columnwidth]{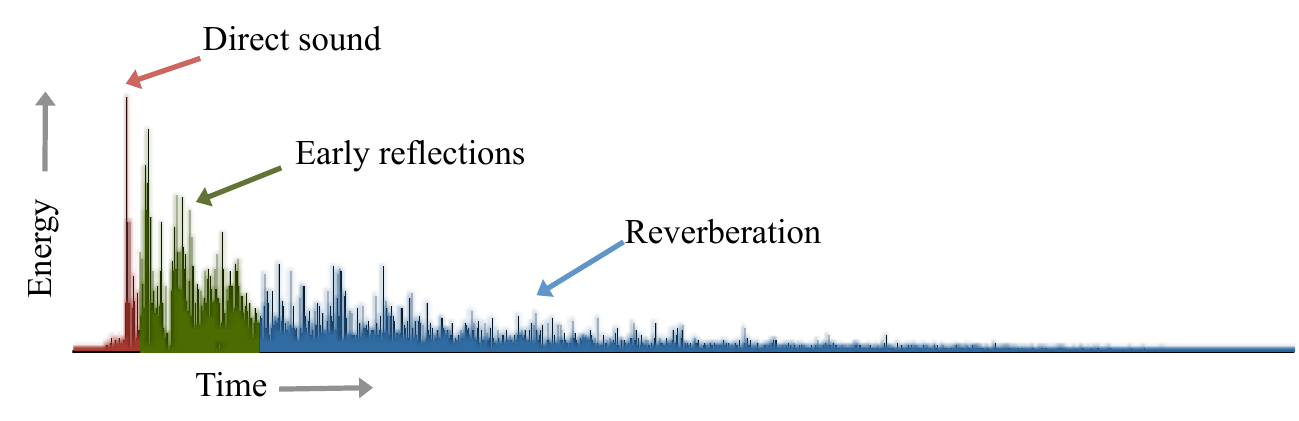}
\caption{The main components of a typical sound propagation impulse response.
The first sound arrival is the direct sound, followed by the early reflections.
Reverberation makes up the remainder of the impulse response.
}\label{fig:ir_parts}
\end{figure}

In indoor environments, the IR can be divided into 3 main components that must be modeled by our sound rendering technique: direct sound, early reflections, and late reverberation.
These are illustrated in Figure~\ref{fig:ir_parts}.
The direct sound is the first sound arrival at the listener's position and is strongly directional.
The early reflections (ER) follow the direct sound and consist of the first few indirect bounces of sound.
Early reflections usually arrive from several distinct directions corresponding to major reflectors in the scene.
The remainder of the impulse response consists of late reverberation (LR).
The late reverberation represents the buildup of many high-order reflections as the sound decays to zero amplitude.
Usually, the earliest part of the LR is somewhat directional, then increasingly diffuse toward the later parts as reflections scatter the sound.
The rate of decay of the LR varies with frequency and is impacted by the effects of materials and air absorption.
In outdoor environments, the impulse response structure is similar, though the relative strength and directivity of the components may be different.
For example, there may be greater emphasis on direct sound and ER rather than LR because the open environment allows much of the later indirect sound to escape the scene.

%%%%%%%%%%%%%%%%%%%%%%%%%%%%%%%%%%%%%%%%%%%%%%%%%%%%%%%%%%%%%%%%%%%%%%%%%%%%%%%%

%\begin{small}
%\bibliographystyle{acmsiggraph}
% \bibliographystyle{ACM-Reference-Format}
% \bibliography{paper}
%\end{small}

%\clearpage
%\newpage
%\pagenumbering{arabic}
%\input{a-appendix}

% \end{document}

\end{document}